\documentclass[a4paper, 11pt, onecolumn, nofootinbib,superscriptaddress]{revtex4}
\usepackage{amsfonts, amssymb, graphicx, float, appendix, multirow, longtable}
\usepackage{amsmath}
\usepackage{color}
\usepackage{hyperref}
\usepackage{mathrsfs}
\usepackage[left=2.5cm,top=1.5cm,right=2.5cm,bottom=3cm]{geometry}
\usepackage{subfigure}

\newcommand{\sfrac}[2]{{\textstyle{#1\over#2}}}
\newcommand{\dd}{\text{D}}
\newcommand{\dl}{\text{d}}
\newcommand{\re}{\text{Re}}
\newcommand{\dett}{\text{det}}
\newcommand{\pp}{\partial}

\usepackage{times}
\begin{document}
\title{\textbf{Characteristic Formulation for Metric $f(R)$ Gravity}}

\author{Bishop Mongwane}
\email{bishop.mongwane@uct.ac.za} 
\affiliation{Department of Mathematics \& Applied Mathematics,
  University of Cape Town, 7701 Rondebosch, South Africa}

\begin{abstract}
In recent years, the Characteristic formulation of numerical relativity has found increasing use in the extraction of gravitational radiation from numerically generated spacetimes. In this paper, we formulate the Characteristic initial value problem for $f(R)$ gravity. We consider, in particular, the vacuum field equations of Metric $f(R)$ gravity in the Jordan frame, without utilising the dynamical equivalence with scalar-tensor theories. We present the full hierarchy of non-linear hypersurface and evolution equations necessary for numerical implementation in both tensorial and \textit{eth} forms. Furthermore, we specialise the resulting equations to situations where the spacetime is almost Minkowski and almost Schwarszchild using standard linearization techniques. We obtain analytic solutions for the dominant $\ell=2$ mode and show that they satisfy the concomitant constraints. These results are ideally suited as testbed solutions for numerical codes. Finally, we point out that the Characteristic formulation can be used as a complementary analytic tool to the $1+1+2$ semi-tetrad formulation.
\end{abstract}

\maketitle
\section{Introduction}

Initial value formulations have a long and eventful history in numerical relativity, dating back to the seminal works of \cite{Arnowitt:1962hi,1979sgrr.work...83Y,Bondi21,Sachs103,hyperboloidal}. This topic has been a subject of several review articles, see for example \cite{0264-9381-25-9-093001} and references therein. For the purposes of fixing context, we recall that relativistic initial value formulations generally come in different flavors, among which, those that are based on a $3+1$ foliation of spacetime are the most popular. The other formulations are Generalised Harmonic, Characteristic and Hyperboloidal. The Generalised Harmonic formulation is based on a harmonic decomposition of the Ricci tensor, resulting in evolution equations for the $4$-metric in some harmonic coordinates \cite{Pretorius:2004jg,Lindblom:2005qh}. The Characteristic approach \cite{Bondi21,Sachs103} is based on foliations of spacetime on outgoing null hypersurfaces while the Hyperboloidal formulation is based on spacetime foliations by spacelike hypersurfaces that smoothly intersect null infinity $\mathscr{I}^{+}$ \cite{hyperboloidal,lrr-2004-1}. In this work, we are interested in setting up a Characteristic formulation of the field equations of metric $f(R)$ gravity.

Geometrically, foliating spacetime with null hypersurfaces presents a natural approach to study gravitational radiation, since these represent the characteristic surfaces of the field equations. Indeed, a Characteristic formulation of the field equations presents a gauge invariant and unambiguous description of gravitational waves in a non-linear setting, where the perturbative methods of 3+1 formulations are not adequate. However, one of the major challenges of characteristic evolutions is the possible development of caustics during evolution. These are coordinate singularities that arise due to focusing of light rays generating the null hypersurfaces. Algorithms to handle this undesirable feature have been proposed \cite{Stewart427,Friedrich345} but there has, apparently, not been a numerical implementation in wide use. Nevertheless, caustic formation is only an issue in standalone evolutions of non-linear spacetimes by characteristic methods. More recent applications of Characteristic formulations are in Cauchy Characteristic Extraction (CCE) and Cauchy Characteristic Matching (CCM) methods. In CCE, one takes metric data on some inner timelike worldtube $\Gamma$, computed from a 3+1 Cauchy code and propagate it to future null infinity $\mathscr{I}^{+}$ via a Characteristic code, thus enabling waveform extraction at $\mathscr{I}^{+}$ \cite{PhysRevD.54.6153,lrr:bishop}. This scheme represents a special case of the more general CCM \cite{1992anr..conf...20B,0264-9381-10-2-015} which, in turn, uses data from the Characteristic code as exact boundary conditions for the metric functions of the 3+1 Cauchy code.

Within the numerical relativity community, there are now a number of Characteristic codes being used, with differing levels of sophistication. For instance, some codes employ second order finite difference schemes \cite{Gomez:2007cj}, others use higher order schemes \cite{Reisswig:2012ka} while others have adopted Spectral methods \cite{Handmer:2014qha}. Another point of distinction among different codes is the coordinate system used to cover the sphere labelling the null directions of the light cones. Common choices range from stereographic coordinate system \cite{Bishop:1997ik} to multi-patch coordinate systems \cite{Reisswig:2006nt,Gomez:2007cj}. There has also been efforts to introduce Adaptive Mesh Refinement schemes to Characteristic evolution codes \cite{Pretorius:2003wc,Thornburg:2009mw}. Overall, these codes have made it possible to demonstrate the versatility of Characteristic methods in numerical relativity and have found extensive applications in, for example, binary black hole mergers \cite{Bishop:1997ik,Reisswig:2009us,Babiuc:2010ze,Handmer:2015dsa}, stellar core collapse \cite{Siebel:2003sp,Reisswig:2010cd,Ott:2010gv}, Einstein-Klein-Gordon systems \cite{Papadopoulos:1996pr,PhysRevD.54.4719,Barreto:2004fn}, Observational Cosmology \cite{vanderWalt:2010zd,vanderWalt:2011jt,Bester:2013fya} etc. These systems represent potential astrophysical laboratories for testing general relativity in the non-linear regime.

Over the years, the theory of general relativity has been subjected to a wide range of experimental tests and has no doubt emerged as one of the most successful theories in Physics. However, there has been considerable interest in the literature to study gravity theories whose Lagrangians contain higher order curvature invariants such as $R^{2}$, $R^{\mu\nu}R_{\mu \nu}$, $R^{\mu\nu\alpha\beta}R_{\mu\nu\alpha\beta}$, $R\Box R$, $R\Box^{k} R$ \cite{Schmidt:2006jt,Sotiriou:2008rp,DeFelice:2010aj}. The motivation for these alternative theories of gravity stems from a variety of grounds, most notably from within the dark sector in Cosmology \cite{Clifton:2011jh}. Moreover, the inflationary paradigm arises naturally in alternative theories of gravity without postulating additional inflaton fields \cite{STAROBINSKY198099,Sotiriou:2008rp,DeFelice:2010aj}. These higher order corrections also arise in the effective action of quantum gravity. For example, in the low energy limit of string theory or when considering compactifications of extra dimensions in M theory \cite{gasperini:1993}. In this work, we restrict our attention to the fourth order metric $f(R)$ gravity. Although simpler than most other alternative theories, general predictions in the theory demands a numerical treatment, especially when considering strong field sources as in numerical relativity.

We derive the full set of non-linear equations necessary for a numerical implementation. We further present linearised solutions about some fixed background spacetimes that may aid in code development in the form of testbed solutions. These solutions are based on a linearization of the exact equations on Minkowski and Schwarzschild backgrounds using standard techniques. In principle, one could consider other background solutions about which to linearize. However, one must be able to analytically cast the metric of such background solutions in Bondi-Sachs form, which is a non-trivial task for most known solutions \cite{Hobill1987}. For example, a Bondi-Sachs representation of the Kerr solution involves elliptic integrals, which require numerical evaluation \cite{nigel:2006}. The existence and stability conditions for both Minkowski and Schwarzschild spacetimes in the context of $f(R)$ gravity have been studied by several authors, see \cite{PhysRevD.46.1475,WHITT1984176,Faraoni:2004is,Nzioki:2013lca}. Within the Bondi-Sachs framework, linearised perturbations, in the manner considered here, have been studied in general relativity by \cite{Bishop:2004ug,Bishop:2011iu,cedeno:2015ucy,Bishop:2015kay}, and have been used as testbed solutions and in analytic descriptions of binary black holes in circular \cite{Bishop:2011iu,cedeno:2015mha} and eccentric orbits \cite{cedeno:2016cxx}. Different approaches on the subject can be found in \cite{PhysRevD.87.104016,Gomez:1994rg,PhysRevD.54.6153}. 

This paper is structured as follows: we review the field equations of metric $f(R)$ and its equivalence to scalar-tensor theories in \S\ref{sec:metric_fr_gravity}. In \S\ref{sec:bondi_sachs_coordinates}, we present the Bondi-Sachs coordinates. The decomposed field equations in tensorial form are given in \ref{sec:main_equations} and  in \S\ref{sec:eth} we present them in the complementary \textit{eth} formalism which is commonly used in numerical codes. We present linearised equations in \S\ref{sec:linearizations} and their solutions when linearised about Minkowski background in \S\ref{sec:minkowski_bg} and Schwarzschild background in \S\ref{sec:schwarzschild}. Finally we conclude in \S\ref{sec:concluding_remarks}. For convenience, we provide the Christoffel symbols for the Bondi-Sachs metric in Appendix \ref{sec:chr_sym}. Throughout this paper, we use Geometrized units $G=c=1$ and metric signature $(-+++)$.

\section{Metric $f(R)$ gravity}
\label{sec:metric_fr_gravity}

\subsection{Field equations}
The gravitational field equations of metric $f(R)$ theories can be derived starting from a simple generalisation of the Einstein-Hilbert action
\begin{equation}
\label{eq:action_j}
S=\frac{1}{16\pi}\int dx^{4}\left[\sqrt{-g} \,f(R) + 16\pi\mathcal{L}_{mat} \right]\;,
\end{equation}
where $f(R)$ is a general function of the Ricci scalar $R$, $g$ is the determinant of the spacetime metric $g_{ab}$, $\mathcal{L}_{mat}$ is the Lagrangian of matter fields. Varying the action (\ref{eq:action_j}) with respect to the metric $g_{ab}$ and assuming that the connection is the Levi-Civita connection\footnote{Relaxing this assumption, such that the affine connection $\Gamma^{a}_{\phantom{a}bc}$ is independent of the metric $g_{ab}$, is the basis of Palatini $f(R)$ and leads to field equations that are different from those of metric $f(R)$ considered here.}, one obtains the equations of motion
\begin{align}
\label{eq:field_eqs1}
\Sigma_{ab} &= 8\pi T_{ab}
\end{align}
where $T_{ab}$ is the energy momentum tensor of standard matter fields, given in terms of the variational derivative of $\mathcal{L}_{mat}$ as
\begin{equation}
T_{ab} =- \frac{2}{\sqrt{-g}}\frac{\delta (\sqrt{-g}\mathcal{L}_{mat})}{\delta g^{ab}}\;.
\end{equation}
The symmetric tensor $\Sigma_{ab}$ is given by,
\begin{align}
\label{eq:field_eqs}
\Sigma_{ab} &= f' R_{ab}-\sfrac{1}{2}fg_{ab}-\nabla_{a}\nabla_{b}f' +g_{ab}\Box f' \nonumber \\
&=  f' R_{ab}-\sfrac{1}{2}fg_{ab}-f''\nabla_{a}\nabla_{b}R - f'''\nabla_{b}R\nabla_{a}R +g_{ab}(f'''\nabla^{c}R\nabla_{c}R + f'' \Box R)\;,
\end{align}
where $\Box=\nabla_{c}\nabla^{c}$ is the d'Alembertian operator and we use $'$ to denote differentiation with respect to the Ricci scalar $R$. Interestingly, $\Sigma_{ab}$ contains terms involving second derivatives of the Ricci scalar $R$ which translates to fourth derivatives of the metric, hence the characterisation as \textit{fourth order gravity}. Unlike in general relativity, the relation between the Ricci scalar $R$ and the trace $T$ of the energy momentum tensor is no longer algebraic ($R=-8\pi T$), but differential, given as
\begin{equation}
\label{eq:trace_eq}
3\,\Box f' -2f+f'R=8\pi T\;.
\end{equation}
Equation (\ref{eq:trace_eq}) governs the dynamics of the scalar degree of freedom inherent in the theory. As in the $3+1$ formulation \cite{Mongwane:2016qtz}, it is convenient to use the equivalent form for the field equations
\begin{equation}
\label{eq:compact_field_eqs_2}
E_{ab}\equiv\Sigma_{ab} - \kappa^{2} T_{ab} - \frac{1}{3} \;g_{ab}(\Sigma-\kappa^{2} T)=0\;.
\end{equation}
where we have introduced the notation $E_{ab}$ for later convenience.
Finally, we note that in the limit of constant scalar curvature $R=R_{0}$, the trace equation (\ref{eq:trace_eq}) reduces to an algebraic relation $-2f+f'R=8\pi T$ and the field equations (\ref{eq:field_eqs}) become
\begin{equation}
R_{ab}-\frac{1}{2}g_{ab}R+\lambda g_{ab}=8\pi T_{ab}
\end{equation}
where $\lambda=R_{0}/4$ is an effective cosmological constant $\Lambda$.

\subsection{Equivalence with scalar-tensor theories}

It has long been known that metric $f(R)$ gravity theories are dynamically equivalent to special cases of Brans-Dicke scalar-tensor theories \cite{BARROW1988515,0305-4470-16-12-022,1983JMP....24.2793T,Wands:1993uu}. We briefly review this equivalence in the following. Starting from the action (\ref{eq:action_j}), one can introduce a new field $\chi$, and recast (\ref{eq:action_j}) into the equivalent form
\begin{equation}
\label{eq:action_js}
S=\frac{1}{16\pi}\int dx^{4}\left[\sqrt{-g} \,f(\chi) + f'(\chi)(R-\chi)\right]+ \int d^{4}x \mathcal{L}_{mat} \;,
\end{equation}
Varying the new action (\ref{eq:action_js}) with respect to $\chi$ leads to
\begin{align}
f''(\chi)(R-\chi) = 0\;.
\end{align}
Then, provided that $f''(\chi) \neq 0$, the above implies $\chi=R$, and consequently, the action (\ref{eq:action_js}) becomes (\ref{eq:action_j}).
If we further define an auxiliary field $\phi$
\begin{equation}
\phi = f'(\chi)
\end{equation}
and supposing that the relation is invertible, then the action (\ref{eq:action_js}) can be expressed as
\begin{equation}
\label{eq:action_jst}
S=\frac{1}{16\pi}\int dx^{4}\sqrt{-g} \, \left[\chi(\phi) R -V(\phi)\right]+ \int d^{4}x \mathcal{L}_{mat} \;,
\end{equation}
where the potential $V(\phi)$ is given by
\begin{equation}
V(\phi) = \chi(\phi)\phi-f(\chi(\phi))\;.
\end{equation}
The action (\ref{eq:action_jst}) corresponds to the Jordan frame representation of a Brans-Dicke scalar-tensor theory without a kinetic term for the scalar field, i.e. with Brans-Dicke parameter ${\omega_{BD}=0}$. By transforming to the Einstein frame, one can proceed to show that this is conformally equivalent to the Einstein-Hilbert action with a scalar field that couples minimally to the Ricci scalar \cite{1989PhRvD..39.3159M}. This equivalence can be a convenient tool when studying various modified gravity theories. However, one should exercise caution when interpreting results, see for example \cite{Faraoni:2006hx,Kainulainen:2007bt,Bahamonde:2016wmz,Jaime:2010kn,Briscese:2006xu,Capozziello:2006dj,PhysRevLett.102.221101}.

\section{The Bondi-Sachs metric}
\label{sec:bondi_sachs_coordinates}
For the Characteristic initial value problem, we employ coordinates $(u,r,x^{A})$ based on a family of outgoing null hypersurfaces emanating from an inner worldtube $\Gamma$ denoting the inner boundary of the characteristic domain. Within this system, $u=r-t$ is a retarded time coordinate labelling the hypersurfaces, $r$ is a surface area coordinate and $x^{A} (A=2,3)$ \footnote{Here, and in the following, we will generally use uppercase indices for the angular directions. These will run from $2$ to $3$.} are labels for the null rays. Then the Bondi-Sachs metric takes the form
\begin{align}
ds^{2} =& \; g_{\mu \nu}\dl x^{\mu}\dl x^{\nu} \\
\label{eq:bs_metric}
=&-\left(e^{2\beta}\frac{V}{r}-r^{2}h_{AB}U^{A}U^{B} \right)\dl u^{2}-2e^{2\beta}\dl u \dl r -2r^{2}h_{AB}U^{B}\dl u \dl x^{A}+r^{2}h_{AB}\dl x^{A}\dl x^{B} \;.
\end{align}
It is straightforward to compute the contravariant components of the Bondi-Sachs metric. The non-zero components are
\begin{align}
g^{rr}=e^{-2\beta}\frac{V}{r},\qquad g^{rA} = -e^{-2\beta}U^{A}, \qquad g^{ru} = -e^{-2\beta}, \qquad g^{AB}=r^{-2}h^{AB}\;.
\end{align}
The Christoffel symbols for the above metric are given in Appendix \ref{sec:chr_sym}. We note that it is sometimes convenient to use $W$ instead of the more usual Bondi-Sachs variable $V$, where $W:=V-r$. The 2-tensor $h_{AB}$, with $h^{AB}h_{BC}=\delta^{A}_{\phantom{A}C}$, satisfies the determinant condition
\begin{align}
\label{eq:det_condition}
\dett(h_{AB}) = \dett (q_{AB})
\end{align}
where $q_{AB}$ is the unit 2-sphere metric, so that $h_{AB}$ has only two degrees of freedom. By considering the metric of $r=const$ surfaces\footnote{This can be obtained from (\ref{eq:bs_metric}) by setting $dr=0$.}, one identifies $h_{AB}$ as the conformal $2$-metric of surfaces of constant $u$ which foliate the worldtube, $e^{2\beta}V/r$ corresponds to the square of the lapse function while $-U^{A}$ represents the shift vector. In total, the metric (\ref{eq:bs_metric}) contains only six free variables $h_{AB}$, $\beta$, $V$ and $U^{A}$, which are in general, a function of the coordinates. Evolution equations for these Bondi-Sachs variables are derived from the field equations of gravity.

\section{The Field Equations}
\label{sec:field_eqs}
In analogy with the $3+1$ formulation, the field equations within the Bondi-Sachs formalism can be classified into Main and Constraint equations. In the following sections, we present these in turn.

\subsection{Main equations}
\label{sec:main_equations}
The Main equations are further classified into hypersurface and evolution equations. The hypersurface equations form a hierarchical set of equations for the Bondi-Sachs variables $\beta$, $U^{A}$ and $V$ to be integrated radially once $h_{AB}$ and $R$ are given on some $u=const$ slice. These are derived from the $R^{u}_{\phantom{u}\alpha}$ components of the field equations, giving

\begin{eqnarray}
\beta_{,r} \left(1+\frac{r}{2}\frac{f''}{f'}R_{,r} \right)&=& \frac{r}{16}h^{AC}h^{BD}h_{AB,r}h_{CD,r} +\frac{r}{4f'}\biggl(f''R_{,rr}+f'''R_{,r}R_{,r} \biggr)
\end{eqnarray}
\begin{eqnarray}
\label{eq:qeq}
(r^{2}Q_{A})_{,r} &=&  2r^{4}\left(r^{-2}\beta_{,A} \right)_{,r} -r^{2}h^{BC}\dd_{C}h_{AB,r} +\frac{2r^{2}}{f'}\biggl\{r^{2}f''\left(r^{-2}R_{,A} \right)_{,r} \nonumber \\
&&+\left.f'''R_{,A}R_{,r}+f''\beta_{,A}R_{,r}- \frac{r^{2}}{2}f''h_{AB}U^{B}_{\phantom{B},r}R_{,r}-\frac{1}{2}f''h^{DC}h_{AC,r}R_{,D}\right\}
\end{eqnarray}
\begin{eqnarray}
2e^{-2\beta}V_{,r} &=& \mathcal{R}-2\dd^{A}\dd_{A}\beta-2\dd^{A}\beta\dd_{A}\beta +r^{-2}e^{-2\beta}\dd_{A}\left(r^{4}U^{A} \right)_{,r} \nonumber \\
& &-\frac{r^{4}}{2}e^{-4\beta}h_{AB}U^{A}_{\phantom{A},r}U^{B}_{\phantom{B},r}+\frac{e^{-2\beta}}{f'}\biggl\{e^{2\beta}f''\dd^{A}\dd_{A}R +f'''e^{2\beta}\dd^{A}R\dd_{A}R  \nonumber \\
&&\left.  -2f''R_{,u}+2Vf''R_{,r}-r^{2}f''R_{,r}\dd_{C}U^{C}-2rU^{C}f''R_{,C} \right. \nonumber \\ 
&&\left. -\frac{r^{2}}{3}e^{2\beta}f + \frac{2r^{2}}{3}e^{2\beta}f'R \right\}
\end{eqnarray}
where in (\ref{eq:qeq}) we have used the auxiliary quantity $Q_{A}$,
\begin{equation}
Q_{A} = r^{2}e^{-2\beta}h_{AB}U^{B}_{\phantom{B},r}\;.
\end{equation}
To obtain the evolution equation for $h_{AB}$, it suffices to consider the trace-free symmetric part of the angular components of the field equations,
\begin{align}
 m^{A}m^{B}&\left[ r\left(rh_{AB,u} \right)_{,r} -\frac{1}{2}\left(rVh_{AB,r}\right)_{,r} -2e^{\beta}\dd_{A}\dd_{B}e^{\beta} -\frac{1}{2}r^{4}e^{-2\beta}h_{BD}h_{AC}U^{C}_{\phantom{C},r}U^{D}_{\phantom{D},r}\right. \nonumber \\
&\left.  +U^{C}r^{2}\dd_{C}h_{AB,r}+h_{AC}\dd_{B}\left(r^{2}U^{C} \right)_{,r} + \frac{1}{2}r^{2}h_{AB,r}\dd_{C}U^{C}-r^{2}h_{BE}h_{AC,r} \left(\dd^{C}U^{E}-\dd^{E}U^{C} \right)\right. \nonumber \\
&\left. -\frac{1}{f'}\left\{f''e^{2\beta}\dd_{A}\dd_{B}R +f'''e^{2\beta}R_{,A}R_{,B} -\frac{r^{2}}{2}f''h_{AB,r}R_{,u} + \frac{r}{2}Vf''h_{AB,r}R_{,r} \right. \right. \nonumber \\
&\left. \left. -r^{2}f''R_{,r}\dd_{A}U_{B} -\frac{r^{2}}{2}f''h_{AB,u}R_{,r} -\frac{r^{2}}{2}f''U^{C}h_{AB,r}R_{,C}\right\}\right]=0
\end{align}
where $m^{A}$ is a complex dyad such that $h^{AB}=m^{(A}\bar{m}^{B)}$. The trace equation (\ref{eq:trace_eq}) gives the following evolution equation for the quantity $f'$
\begin{align}
&-\frac{2}{r}\pp_{u}\pp_{r}\left(rf' \right)+\frac{V}{r}\pp_{r}\pp_{r}f' +\frac{1}{r}\pp_{r}V\pp_{r}f'+\frac{V}{r^{2}}\pp_{r}f'-2U^{A}\pp_{A}\pp_{r}f'\nonumber \\
&-\pp_{r}f'\dd_{A}U^{A} -\frac{2}{r}U^{A}\pp_{A}f'-\pp_{A}f'\pp_{r}U^{A}+r^{-2}e^{2\beta}\dd_{A}\dd^{A}f'\nonumber \\
&+2e^{2\beta}r^{-2}h^{AC}\pp_{C}f'\pp_{A}\beta = \frac{2}{3}e^{2\beta}f-\frac{1}{3}e^{2\beta}f'R
\end{align}
To turn this into an equation for the Ricci scalar $R$, one uses the fact that $f'= d f(R)/dR$, and proceed via the Chain Rule such that
\begin{subequations}
\label{eq:chain_rule}
\begin{eqnarray}
f'_{\phantom{'},x} &=& f'' R_{,x} \\
f'_{\phantom{'},xy} &=& f'' R_{,xy}+f'''R_{,x}R_{,y}
\end{eqnarray}
\end{subequations}

\subsection{Conservation Conditions}
\label{supplemtary_equations}
up to this point, we have only focused on the main equations. The remaining components of the field equations $R^{r}_{\phantom{r}\alpha}$, split into the Trivial equation 
\begin{equation}
\label{eq:ctrivial}
E_{ur} = 0
\end{equation}
and Supplementary equations
\begin{equation}
\label{eq:csupplementary}
E_{uu} = 0 \qquad \text{and}\qquad E_{Au}=0\;.
\end{equation}
where we have used the notation (cf. Equation (\ref{eq:compact_field_eqs_2}))
\begin{equation}
\label{eq:compact_field_eqs_3}
E_{ab}\equiv\Sigma_{ab} - \kappa^{2} T_{ab} - \frac{1}{3} \;g_{ab}(\Sigma-\kappa^{2} T)=0\;.
\end{equation}
Along with the Main equations, these make up the full set of components for the field equations. Because of the Bianchi identities, and assuming that the Main equations are satisfied, the Trivial equation is satisfied identically, while the supplementary equations need only be satisfied on a single spherical cross-section of the world tube as was shown in the general relativity case by \cite{Bondi21,Sachs103}.

Clearly, a key to this conservation property is the Bianchi identities. In $f(R)$ gravity, the divergence of the field equations takes the form \cite{Koivisto:2005yk}
\begin{eqnarray}
\nabla^{a}\Sigma_{ab}&=&  \nabla^{a}\left(f' R_{ab}-\sfrac{1}{2}fg_{ab}-\nabla_{a}\nabla_{b}f' +g_{ab}\Box f' \right)=0 \nonumber \\
&=& R_{ab}\nabla^{a}f'+f'\nabla^{a} R_{ab}-\sfrac{1}{2}g_{ab}\nabla^{a} f-\nabla^{a}\nabla_{a}\nabla_{b}f' +g_{ab}\nabla^{a}\nabla^{c}\nabla_{c} f'  \\
&=& R_{ab}\nabla^{a}f'+f'\nabla^{a} R_{ab}-\sfrac{1}{2}g_{ab}f'\nabla^{a} R-(\nabla^{a}\nabla_{b}\nabla_{a} -\nabla_{b}\nabla^{c}\nabla_{c}) f'  
\end{eqnarray}
Then the generalised Bianchi identities $\nabla^{a}\Sigma_{ab}=0$ follow geometrically because 
\begin{equation}
\nabla^{a}(R_{ab}-\sfrac{1}{2}g_{ab}R)=0 \qquad \text{and} \qquad (\nabla^{a}\nabla_{b}\nabla_{a} -\nabla_{b}\nabla^{c}\nabla_{c})f'=R_{ab}\nabla^{a}f'
\end{equation}
as a result of the standard Bianchi and Ricci identities.

General expressions for (\ref{eq:ctrivial}) and (\ref{eq:csupplementary}) are lengthy and are not required in most numerical applications. We give instead linearised expressions in \S\ref{sec:linearizations}.

\section{Spin weighted and \textit{eth} formalism}
\label{sec:eth}

Within the spin weighted formalism, the unit sphere metric $q_{AB}$ is expressed in terms of a dyadic product $q_{AB}=q_{(A}\bar{q}_{B)}$, where the dyad $q^{A}$ is a complex\footnote{We will generally use an overbar on a complex quantity to denote complex conjugation.} basis 2-vector satisfying $q^{A}q_{A}=0$, $q^{A}\bar{q}_{A}=2$ and $q_{A}=q_{AB}q^{B}$ \cite{1967JMP.....8.2155G,Gomez:1996ge}. We note that the basis vectors are not unique, up to a phase transformation. For a given $q_{A}$, one can construct an alternative basis $\hat{q}_{A}=e^{i\alpha}q_{A}$, where the phase $\alpha$ is real. Using the dyad vectors $q^{A}$, rank-$n$ tensor fields $T_{A_{1}A_{2}\cdots A_{n}}$ on the sphere can be conveniently represented by scalar fields,
\begin{equation}
\label{eq:scalar_field}
T=q^{A_{1}}\cdots q^{A_{m}}\bar{q}^{A_{m+1}}\cdots \bar{q}^{A_{n}}\,T_{A_{1}\cdots A_{n}}\;.
\end{equation}
The spin-weight $s$ of such scalar fields depends on the rank $n$ of the tensor field and is given by $s=2m-n$, where $m$ is the number of $q^{A}$ factors and $n-m$ represents the number of $\bar{q}^{A}$ factors appearing in (\ref{eq:scalar_field}). In general, the scalars (\ref{eq:scalar_field}) will have the transformation property $T\rightarrow e^{i\alpha s}T$. With this in mind, the three spin-weighted scalars 
\begin{align}
\label{eq:jjbk}
J=\frac{1}{2}q^{A}q^{B}h_{AB}, \qquad \bar{J}=\frac{1}{2}\bar{q}^{A}\bar{q}^{B}h_{AB} \qquad \text{and} \qquad K=\frac{1}{2}q^{A}\bar{q}^{B}h_{AB}
\end{align}
with respective spin weights $+2$, $-2$ and $0$, contains all the degrees of freedom of the $2-$tensor $h_{AB}$. Using (\ref{eq:jjbk}), $h_{AB}$ is irreducibly decomposed as 
\begin{align}
2h_{AB} = \bar{J}q_{A}q_{B}+J\bar{q}_{A}\bar{q}_{B}+K\left(q_{A}\bar{q}_{B}+\bar{q}_{A}q_{B} \right)\;,
\end{align}
with the inverse $2-$metric $h^{AB}$ given by
\begin{align}
2h^{AB} = -\bar{J}q^{A}q^{B}-J\bar{q}^{A}\bar{q}^{B}+K\left(q^{A}\bar{q}^{B}+\bar{q}^{A}q^{B} \right)\;.
\end{align}
Furthermore, the determinant condition (\ref{eq:det_condition}) implies the relation
\begin{equation}
\label{eq:determinant_condition_2}
K^{2}=1+J\bar{J}\;.
\end{equation}
Consequently, the scalar $K$ contains no additional information and $h_{AB}$ is uniquely determined by $J$, for an arbitrary Bondi-Sachs metric. Similarly, $U^{A}$ and $Q^{A}$ are decomposed into the spin-weighted fields
\begin{equation}
U=U^{A}q_{A} \qquad \bar{U}=U^{A}\bar{q}_{A} \qquad Q=Q^{A}q_{A} \qquad \bar{Q}=Q^{A}\bar{q}_{A}
\end{equation}
with respective spins of $+1$, $-1$, $+1$ and $-1$. We note that within this spin weighted formalism, the scalar quantities $\beta$, $V$ and $R$ are spin-$0$ fields.

In addition to the spin weighted scalars, it is convenient to define complex differential eth operators $\eth$ and $\bar{\eth}$ whose action on a quantity $X$ of spin weight $s$ is given as
\begin{equation}
\eth X = q^{A}\pp_{A}X+s\Upsilon X\;, \qquad \bar{\eth} X = \bar{q}^{A}\pp_{A}X-s\bar{\Upsilon} X\;.
\end{equation}
where 
\begin{equation}
\label{eq:ups}
\Upsilon = -\frac{1}{2} q^{A}\bar{q}^{B}\nabla_{A}q_{B}\;.
\end{equation}
The resulting quantities $\eth X$ and $\bar{\eth} X$ have spin weights $s+1$ and $s-1$, respectively. More generally, the operator $\eth$ ($\bar{\eth}$) acting on a spin weighted scalar has the effect of raising (lowering) the spin weight by 1.

For the stereographic coordinate system $x^{A}=(q,p)$, which we adopt in this work, the unit sphere metric $q_{AB}$ is given as
\begin{equation}
q_{AB}dx^{A}dx^{B}=\frac{4}{q^{2}+p^{2}+1}\left(dq^{2}+dp^{2}\right)\;.
\end{equation}
The dyad vectors then become
\begin{equation}
q^{A}=\frac{q^{2}+p^{2}+1}{2}(1,i) \qquad \text{and} \qquad q_{A}=\frac{2}{q^{2}+p^{2}+1}(1,i)\;.
\end{equation}
With this choice, (\ref{eq:ups}) becomes $\Upsilon=q+ip$.

Using the above formalism, the Hypersurface equations become
\begin{subequations}
\label{eq:nnl_eqs}
\begin{eqnarray}
\beta_{,r} \left(1+\frac{r}{2}\frac{f''}{f'}R_{,r} \right) &=& N_{\beta} + M_{\beta}\\
U_{,r} &=& r^{-2}e^{2\beta}Q+N_{U} \\
(r^{2}Q)_{,r} &=& -r^{2}\left(\bar{\eth}J+\eth K \right)_{,r}+2r^{4}\eth\left(r^{-2}\beta \right)_{,r} + N_{Q}+M_{Q}\\
W_{,r} &=& \frac{1}{2}e^{2\beta}\mathcal{R} -1 - e^{\beta}\eth \bar{\eth}e^{\beta}+\frac{1}{4}r^{-2}\left[r^{4}\left(\eth\bar{U}+\bar{\eth}U \right) \right]_{,r}\nonumber \\
&& +N_{W} + \frac{1}{2}e^{2\beta} M_{W}
\end{eqnarray}
\end{subequations}
where the 2--Ricci scalar $\mathcal{R}$ is given by
\begin{equation}
\mathcal{R} = 2K-\eth\bar{\eth}K+\frac{1}{2}\left(\bar{\eth}^{2}J+\eth^{2}\bar{J} \right)+\frac{1}{4K}\left(\bar{\eth}\bar{J}\eth J - \bar{\eth}J \eth\bar{J} \right)
\end{equation}
The evolution equations become
\begin{eqnarray}
2\left(rJ \right)_{,ur} &=& \left[r^{-1}V(rJ)_{,r} \right]_{,r} -r^{-1}\left(r^{2}\eth U \right)_{,r}+2r^{-1}e^{\beta}\eth^{2}e^{\beta}\nonumber \\
&&-J\left(r^{-1}W \right)_{,r}+N_{J}+r^{-1}M_{J}
\end{eqnarray}

\begin{eqnarray}
2\left(rf' \right)_{,ur} &=& \left[r^{-1}V(rf')_{,r} \right]_{,r} -f'\left(r^{-1}W \right)_{,r}-U\bar{\eth}f'-\bar{U}\eth f'\\
&& +\frac{r^{-1}e^{2\beta}}{K}\left[J \left(\eth \bar{J} \bar{\eth}f' + \bar{\eth}\bar{J}\eth f'\right) +\bar{J} \left( \bar{\eth}J\eth f' +\eth J\bar{\eth}f'\right) \right] \nonumber \\
&& +Ke^{2\beta}r^{-1}\left(\eth \bar{\eth} f'+\bar{\eth}f'\eth \beta +\eth f'\bar{\eth}\beta \right)-\frac{r}{2}\left[\bar{\eth}f'U_{,r}+\eth f'\bar{U}_{,r} \right. \nonumber \\
&& \left. + f'_{,r}(\eth \bar{U}+\bar{\eth}U) + 2(\bar{U}\eth f'_{,r}-U\bar{\eth}f'_{,r})\right]-\frac{r^{-1}e^{-2\beta}}{2}\left[\bar{J}\eth^{2}f'+J\bar{\eth}^{2}f' \right. \nonumber \\
&& \left. +\bar{\eth}J\bar{\eth}f' +\eth \bar{J}\eth f'+2\left(J\bar{\eth}f'\bar{\eth}\beta +\bar{J}\eth f'\eth \beta  \right) \right]
 + \frac{re^{2\beta}}{3}\left(2f-f'R\right)
\end{eqnarray}
where, again, one is to use the Chain Rule (\ref{eq:chain_rule}) to obtain an evolution equation for the Ricci scalar $R$. The terms $N_{\beta}$, $N_{U}$, $N_{Q}$, $N_{W}$ and $N_{J}$ are non-linear aspherical terms whose representation in terms of spin-weighted variables is given in \cite{Bishop:1997ik}. The terms $M_{\beta}$, $M_{Q}$, $M_{W}$ and $M_{J}$ are modified gravity terms arising from the $f(R)$ corrections. These can be computed as 
\begin{eqnarray}
M_{\beta} &=& \frac{r}{4f'}\biggl(f''R_{,rr}+f'''R_{,r}R_{,r} \biggr)\;,\\
f'M_{Q} &=& r\left(f''r^{-1}\eth R\right)_{,r} -\frac{1}{2}r^{2}e^{-2\beta}f''R_{,r}\left(KU_{,r}+J\bar{U}_{,r} \right)-f''R_{,r}\eth \beta \nonumber \\
&& -\frac{1}{2}f''K\left(K_{,r}\eth R +J_{,r}\bar{\eth}R \right)+\frac{1}{2}f''\left( \bar{J}J_{,r}\eth R + JK_{,r}\bar{\eth}R \right) \;,\\
f'M_{W} &=& -\frac{1}{2}r^{2}f''e^{-2\beta}R_{,r}\left(\eth \bar{U}+\bar{\eth}U \right) - rf''e^{-2\beta}\left(\bar{U}\eth R + U\bar{\eth}R \right)-\frac{1}{2}f''\left[\bar{\eth}\left(J\bar{\eth}R\right) + \eth \left(\bar{J} \eth R\right) \right] \nonumber \\
&&-\frac{1}{2}f'''\left[\bar{J}\left(\eth R \right)^{2}-2K\bar{\eth}R\eth R +J\left(\bar{\eth}R \right)^2 \right]+2f''e^{-2\beta}\left( R_{,r}V -rR_{,u}\right)\nonumber \\
&&+\frac{1}{2}f''\left(\eth R \bar{\eth}K+\bar{\eth}R\eth K \right)+f''K\bar{\eth}\eth R -\frac{r^{2}}{3}\left(f-2f'R\right)\;,\\ 
f'M_{J} &=& f''\eth\eth R + f'''(\eth R)^2 -\frac{1}{2}f''(J\eth\bar{J}\eth R +J\bar{\eth}J\bar{\eth}R+K\eth J \bar{\eth} R-K\bar{\eth}J\eth R-2J\eth K\bar{\eth}R) \nonumber \\
&&-\frac{1}{2}e^{-2\beta}f''\left(r^{2}J\right)_{,r}\left[\bar{U}\eth R + U\bar{\eth}R \right] +f''e^{-2\beta}R_{,r}Vr^{-1}\left(r^{2}J\right)_{,r} \nonumber \\
&&-\frac{1}{2}f''e^{-2\beta}r^{2}R_{,r}\left(2K\eth U + 2J\eth \bar{U} +U\bar{\eth}J +\bar{U}\eth{J}\right)-f''e^{-2\beta}R_{,u}\left(r^{2}J\right)_{,r} \nonumber \\
&&-r^{2}f''e^{-2\beta}R_{,r}J_{,u} \;.
\end{eqnarray}
As in the $3+1$ case, it may be necessary to define $\psi=R_{,u}$ so that the Hypersurface equations contain no $u$ derivatives.

\section{Linearised perturbations}
\label{sec:linearizations}

In the following, we specialise the above non-linear equations to situations where the spacetime is almost Schwarzschild and almost Minkowski.
In outgoing null coordinates, the Schwarzschild metric takes the Eddington-Finkelstein form
\begin{align}
\label{eq:background}
ds^{2} = - \left(1-\frac{2M}{r} \right)\dl u^{2}-2\dl u\dl r + r^{2}q_{AB}\dl x^{A}\dl x^{B}\;,
\end{align}
where it is to be understood that $M=0$ corresponds to Minkowski space. The existence and stability of both Schwarzschild and Minkowski spacetimes in $f(R)$ gravity can be found in, for example, \cite{PhysRevD.46.1475,WHITT1984176,Faraoni:2004is,Nzioki:2013lca}. The line element (\ref{eq:background}) corresponds to $J=U=\beta=0$ and $W=-2M$. We therefore designate the following quantities and their derivatives as first order,
\begin{equation}
\label{key_vars}
J,\bar{J},U,\bar{U},w,\beta = \mathcal{O}(\epsilon)
\end{equation}
with $W=-2M+w$. We note that the scalar $K$ is unity to linear order because of the determinant condition (\ref{eq:determinant_condition_2}). The linearization procedure proceeds by discarding terms of order $\mathcal{O}(\epsilon^{2})$ and higher, i.e. terms involving products of the first order quantities (\ref{key_vars}). We note that the Ricci scalar $R$ vanishes for the background metric (\ref{eq:background}). In order to deal with the $f(R)$ corrections, we therefore perform a Taylor expansion about the background such that, to linear order\footnote{We use the fact that $f(0)=0$, which is one of the conditions for the stability of the Schwarzschild solution in $f(R)$ gravity \cite{Nzioki:2013lca}.},
\begin{align}
f(R) = f'_{(0)}R\;.
\end{align}
where $f'_{(0)}$ is a background quantity and $R=\mathcal{O}(\epsilon)$. To avoid having to write pre-factors $f'_{(0)}$ and $f''_{(0)}$, we note that one can define an effective mass for the Scalaron field as
\begin{equation}
m^{2} = \frac{1}{3}\frac{f'}{f''}\;.
\end{equation}
The linearised Main equations (\ref{eq:nnl_eqs}) then become
\begin{subequations}
\label{eq:linear_eqs}
\begin{eqnarray}
\label{eq:lbeta}
\beta_{,r} - \frac{1}{3m^{2}}R_{,rr}=0\;,
\end{eqnarray}
\begin{eqnarray}
r^{3}U_{,rr}+4r^{2}U_{,r}+r\bar{\eth}J_{,r}+4\eth \beta - 2r\eth \beta_{,r} +\frac{2}{3m^{2}}R-\frac{2r}{3m^{2}}R_{,r}=0\;,
\end{eqnarray}
\begin{eqnarray}
4\beta - 2\eth\bar{\eth}\beta +\frac{1}{2}\left(\bar{\eth}^{2}J+\eth^{2}J \right)+\frac{1}{2r^{2}}\left[r^{4}\left(\eth\bar{U}+\bar{\eth}U \right) \right]_{,r} -2w_{,r} -\frac{r^{2}}{3}R \nonumber \\
-\frac{2r}{3m^{2}}\left(1-\frac{2M}{r} \right)R_{,r}+\frac{2}{3m^{2}} R_{,u} =0\;,
\end{eqnarray}
\begin{eqnarray}
2r\left(rJ \right)_{,ur} -2\eth^{2}\beta + 2r\eth U+r^{2}\eth U_{,r}-2(r-M)J_{,r} \nonumber \\
-r^{2}\left(1-\frac{2M}{r} \right)J_{,rr}-\frac{1}{3m^{2}}\eth\eth R= 0 \;,
\end{eqnarray}
\begin{eqnarray}
\label{eq:ltrace}
 \left(1-\frac{2M}{r}\right)R_{,rr}-\frac{2}{r} (R_{,u}+rR_{,ur}) +\frac{2}{r}\left(1-\frac{M}{r} \right)R_{,r}+r^{-2}\eth\bar{\eth}R-m^{2}R=0\;.
\end{eqnarray}
\end{subequations}
A noteworthy feature of the above equations is that the $f(R)$ terms have pre-factors of $1/m^{2}$. Therefore as $m\rightarrow \infty$, the equations will resemble those of general relativity. This is the basic principle behind screening mechanisms that allow modified gravity to behave like general relativity in certain environments by suitably altering the mass of the Scalaron field.

The trivial equation (\ref{eq:ctrivial}) simplifies to
\begin{eqnarray}
\label{eq:ltrivial}
&& \frac{1}{r^{2}}\left [2\left(r-M \right)\beta_{,r}+r^{2}\left(1-\frac{2M}{r} \right)\beta_{,rr} +\frac{1}{2}rw_{,rr} +\eth\bar{\eth}\beta \right. \nonumber \\
&&\left. -2r^{2}\beta_{,ru}- \frac{1}{4}\left[r^{2} \left(\eth \bar{U}+\bar{\eth} U \right) \right]_{,r}\right] = \frac{1}{3m^{2}} \left(R_{,ur}-\frac{M}{r^{2}}R_{,r}\right)-\frac{1}{6}R  \;,
\end{eqnarray}
while the constraints (\ref{eq:csupplementary}) respectively become 
\begin{align}
\label{eq:uAc}
& \frac{1}{4r^{2}} \left[-4r^{2}\eth \beta_{,u} +2r^{2}\bar{\eth} J_{,u}-2r^{4}U_{,ur} + 4r^{2}U+ 2r\eth w_{,r}-2\eth w \right. \nonumber \\
& \left.+r^{2}\left(\eth\bar{\eth}U-\eth\eth \bar{U}\right) +2r^{2}\left(r-2M \right)\left(4U_{,r}+rU_{,rr}\right)\right] = \frac{1}{3m^{2}}\eth R_{,u}\;,
\end{align}
and
\begin{align}
\label{eq:uuc}
&\frac{1}{2r^{3}}\left [-4r(r-2M)\beta_{,u} +2(r-2M)\eth\bar{\eth}\beta +r(r-2M)w_{,rr} +\eth\bar{\eth}w+2rw_{,u}\right. \nonumber \\
&\left. -Mr(\eth \bar{U}+\bar{\eth} U)-r^{3}\left( \eth \bar{U}+\bar{\eth} U\right)_{,u} -4r^{2}(r-2M)\beta_{,ru}  +2r \left(r-2M \right)^{2}\beta_{,rr} \right. \nonumber \\
& \left.+4\left(r-2M\right)\left(r-M \right)\beta_{,r}\right] = \frac{1}{3m^{2}}\left[R_{,uu} + \frac{M}{r^{2}}R_{,u} - \frac{M}{r^{2}} \left(1-\frac{2M}{r} \right)R_{,r} \right] \nonumber \\
& -\frac{1}{6}\left(1-\frac{2M}{r}\right)R\;.
\end{align}
Finally, one can derive an expression for the linearised Ricci scalar $R$ from the metric variables. One is free to do so since in metric $f(R)$ gravity, one assumes that the Christoffel symbols are related to derivatives of the metric in the usual way, unlike in Palatini $f(R)$ gravity. Therefore, from $R=g^{ab}R_{ab}$ one obtains
\begin{eqnarray}
\label{eq:ricci}
R &=& -\frac{4}{r^{2}}\eth \bar{\eth}\beta - \frac{4M}{r^{2}}\beta_{,r}+\frac{4}{r^{2}}\beta-2\left(1-\frac{2M}{r} \right)\beta_{,rr} +4\beta_{,ur} \nonumber \\
&& - \frac{1}{r^{3}}\left(r^{2}w_{,r} \right)_{,r} + \frac{1}{2r^{2}}\left(\bar{\eth}^{2}J+\eth^{2}\bar{J} \right)+\frac{1}{r^{3}}\left(r^{3}\eth\bar{U}+r^{3}\bar{\eth}U \right)_{,r}\;.
\end{eqnarray}
This expression can be used as a consistency check with the result obtained by integrating the trace equation \ref{eq:ltrace}.

\subsection{Eigenfunction decomposition}

It is convenient to write the metric quantities in terms of eigenfunctions of the $\eth$ and $\bar{\eth}$ operators. Without loss of generality, we assume that the linearised variables can be written as \cite{Bishop:2004ug},
\begin{subequations}
\label{eq:igenfunction_decomposition}
\begin{eqnarray}
R &=& R_{0}(r)\,\re\left(e^{i\nu u} \right) Z_{\ell m}\\
\beta &=& \beta_{0}(r)\,\re\left(e^{i\nu u} \right) Z_{\ell m}\\
w &=& w_{0}(r)\,\re\left(e^{i\nu u} \right) Z_{\ell m} \\
U &=& U_{0}(r)\,\re\left(e^{i\nu u} \right) \eth Z_{\ell m} \\
J &=& J_{0}(r)\,\re\left(e^{i\nu u} \right) \eth^{2}Z_{\ell m} 
\end{eqnarray}
\end{subequations}
A more consistent representation would be in terms of a multipolar series involving sums over $\ell$ and $m$ as is done in, for example, \cite{cedeno:2015mha}. The above corresponds to having these quantities fixed, which is sufficient for our purposes. In (\ref{eq:igenfunction_decomposition}) the ${}_{s}Z_{\ell m}$ are orthonormal real-valued spin $s$ spherical harmonics defined as \cite{1967JMP.....8.2155G}
\begin{equation}
{}_{s}Z_{\ell m} = \begin{cases}
\dfrac{i}{\sqrt{2}}\left[(-1)^{m}{}_{s}Y_{\ell m} +{}_{s}Y_{\ell\; -m} \right] & \text{ for } m < 0\\
\hfill {}_{s}Y_{\ell m} \hfill & \text{ for } m=0\\
\dfrac{1}{\sqrt{2}}\left[(-1)^{m}{}_{s}Y_{\ell m} +{}_{s}Y_{\ell\; -m} \right] & \text{ for } m > 0
\end{cases}
\end{equation}
The ${}_{s}Y_{\ell m}$ are the standard spin-weighted spherical harmonics 
\begin{equation}
{}_{s}Y_{\ell m} = \begin{cases}
\sqrt{\dfrac{(\ell - s)!}{(\ell+s)!}}\eth^{s}Y_{\ell m} & \text{ for } s\geq 0 \\\\
(-1)^{s}\sqrt{\dfrac{(\ell - s)!}{(\ell+s)!}}\eth^{-s}Y_{\ell m} & \text{ for } s< 0
\end{cases}
\end{equation}

\subsection{Master equation}
Using the ansatz (\ref{eq:igenfunction_decomposition}), we are able to reduce the linearised equations (\ref{eq:linear_eqs}) into a set of linear ordinary differential equations in $r$, for the quantities $\beta_{0}$, $U_{0}$, $w_{0}$, $J_{0}$ and $R_{0}$. For brevity, we shall henceforth drop the zero subscript on these quantities. In the following, we restrict our attention to the particular case of $\ell=2$. We emphasise that this choice is motivated by simplicity; it is possible to consider other $\ell$ values. We further make the change of variable $r=1/x$. With these simplifications, the linearised equations become

\begin{subequations}
\begin{eqnarray}
\label{eq:bode}
4x\beta_{,x}+\frac{x^{2}}{3m^{2}}R_{,xx}+\frac{2x}{3m^{2}}R_{,x}=0\;,
\end{eqnarray}
\begin{eqnarray}
\label{eq:uode}
4\beta+2x\beta_{,x}+xU_{,xx}-2U_{,x}+4xJ_{,x}+ \frac{2}{3m^{2}}R +\frac{2x}{3m^{2}}R_{,x} = 0\;,
\end{eqnarray}
\begin{eqnarray}
\label{eq:wode}
16x\beta +24xJ-24U+6xU_{,x}+2x^{3}w_{,x} - \frac{x^{-2}}{3}R + \frac{2}{3m^{2}}(i\nu + 3x)R \nonumber \\
+\frac{2x^{2}}{3m^{2}}(1-2xM)R_{,x}=0\;,
\end{eqnarray}
\begin{eqnarray}
-4x\beta +4U -2xU_{,x}+4x^{3}MJ_{,x}-2x^{3}(1-2xM)J_{,xx}\nonumber \\
+4i\nu J -4xi\nu J_{,x}-\frac{2x}{3m^{2}}R =0\;,
\end{eqnarray}
\begin{eqnarray}
\label{eq:rode}
x^{4}(1-2xM)R_{,xx}-2x^{2}(x^{2}M-i\nu)R_{,x}-\left(2xi\nu-6x^{2}+m^{2}\right)R=0\;.
\end{eqnarray}
\end{subequations}
Using standard techniques, it is possible to derive a master equation for the Bondi-Sachs variable $J$. Interestingly, this takes the same form as that obtained in the general relativity case \cite{Bishop:2004ug}.
\begin{align}
\label{eq:jode}
x^{3}(1-2xM)J_{,xxxx}+(4x^{2}+2 i\nu x -14x^{3}M)J_{,xxx}-(4x+16Mx^{2}+2i\nu)J_{,xx} = 0\;.
\end{align}
This master equation can be further simplified by defining an auxiliary variable $J_{,xx}=J_{2}$ \cite{Bishop:2004ug}. Then $J_{2}$ obeys
\begin{align}
\label{eq:jode}
x^{3}(1-2xM)J_{2,xx}+(4x^{2}+2 i\nu x -14x^{3}M)J_{2,x}-(4x+16Mx^{2}+2i\nu)J_{2} = 0\;.
\end{align}
We are now in a position to solve the above linearised ordinary differential equations for the various metric quantities.

\subsection{Solutions}
The solution procedure proceeds in a hierarchical order, mirroring that of a numerical scheme. First we obtain solutions for $J$ and $R$ from (\ref{eq:jode}) and (\ref{eq:rode}). Having obtained $R$, (\ref{eq:bode}) can be solved for $\beta$. Having $\beta$, $R$ and $J$, Equation (\ref{eq:uode}) can be solved for $U$, and finally Equation (\ref{eq:wode}) is solved for $w$. In the following sections, we consider separately the cases of Minkowski (${M=0}$) and Schwarzschild ($M\neq 0$) backgrounds. In all cases, we verify that the $R$ obtained by solving (\ref{eq:rode}) is consistent with that reconstructed from (\ref{eq:ricci}). We also evaluate the constraints by plugging in the obtained solutions.

\subsubsection{Minkowski background}
\label{sec:minkowski_bg}

Following the above procedure, we first consider the static case, $\nu=0$, obtaining the solutions
\begin{align}
R  =& C_{1}x e^{m/x}\left(m^2 -3xm+3x^{2}\right) + C_{2}xe^{-m/x}\left(m^2 +3xm+3x^{2}\right) \displaybreak[1]\\
\beta  =& \frac{C_{1}}{12m^2}e^{\frac{m}{x}}\left(12x^{2}m-5xm^2-12x^3+m^3 \right)\nonumber \\
& -\frac{C_{2}}{12m^2}e^{-\frac{m}{x}}\left(12x^{2}m+5xm^2+12x^3+m^3 \right) + C_{3}\displaybreak[1]\\
J  =& C_{4}+\frac{C_{5}}{x^2}+C_{6}x+C_{7}x^{3}\displaybreak[1]\\
U  =& \frac{x}{6m^2}R +\frac{2C_{5}}{x}+2x^{2}C_{6}+2xC_{3}-3x^{4}C_{7}\displaybreak[1]\\
w  =& -\frac{C_{1}}{6m^{2}x}e^{\frac{m}{x}}\left(6x^2m-6x^{3}+m^3-3m^2x \right) -6x^{2}C_{7}-\frac{10}{x}C_{3}+\frac{12}{x}C_{4}\nonumber\\
& + \frac{C_{2}}{6m^{2}x}e^{-\frac{m}{x}}\left(6x^{2}m+6x^{3}+m^{3}+3m^{2}x \right) -\frac{6}{x^{3}}C_{5}+C_{8}
\end{align}
As expected, the trivial equation (\ref{eq:ltrivial}) is identically satisfied. The constraints (\ref{eq:uAc}) and (\ref{eq:uuc}) respectively lead to
\begin{align}
C_{8} &= 0 \;,\\
4\left(2C_{3}-3C_{4} \right)-C_{8}x &= 0 \;.
\end{align}
For the dynamic case, $\nu \neq 0$, we obtain
\begin{align}
R  =& iC_{1}x \exp\left(\frac{i\nu-\sqrt{m^{2}-\nu^{2}}}{x}\right) \left(m^2-\nu^{2} +3x\sqrt{m^{2}-\nu^{2}}+3x^{2}\right) \nonumber \\
&+ iC_{2}x\exp\left(\frac{i\nu+\sqrt{m^{2}-\nu^{2}}}{x}\right)\left(m^2-\nu^{2} -3x\sqrt{m^{2}-\nu^{2}}+3x^{2}\right)\displaybreak[1]\\
\beta  =& -\frac{C_{1}}{12m^2}\exp\left(\frac{i\nu-\sqrt{m^{2}-\nu^{2}}}{x}\right)\left[5ix(m^{2}-\nu^{2}) +3x\left(4ix+\nu\right)\left(x+\sqrt{m^{2}-\nu^{2}} \right) \right.\nonumber \\
&\left.+\left(m^{2}-\nu^{2} \right)\left(\nu+i\sqrt{m^{2}-\nu^{2}} \right)\right] -\frac{C_{2}}{12m^2}\exp\left(\frac{i\nu+\sqrt{m^{2}-\nu^{2}}}{x}\right)  \biggl[5ix(m^{2}-\nu^{2}) \nonumber \\
&\left. + 3x(4ix+\nu)\left(x-\sqrt{m^{2}-\nu^{2}}\right) +(m^{2}-\nu^{2})(\nu+i\sqrt{m^{2}-\nu^{2}}) \right] + C_{3}\displaybreak[1]\\
J  =& C_{4}+C_{5}x+\frac{C_{6}x^{3}}{6}+\frac{C_{7}}{2}\exp\left(\frac{2i\nu}{x}\right)\left(x-i\nu \right)^{2}x\displaybreak[1]\\
U  =& \frac{x}{6m^2}R -i\nu C_{4}+2C_{5}x^{2}+2xC_{3}-\frac{C_{6}x^{3}}{6}(4i\nu +3x)\nonumber \\
& +\frac{C_{7}x^{3}}{2}\exp\left(\frac{2i\nu}{x} \right)(2i\nu-3x)  \displaybreak[1]\\
w  =& -\frac{C_{1}}{6m^{2}x}\exp\left(\frac{i\nu-\sqrt{m^{2}-\nu^{2}}}{x}\right) \left[(m^{2}-\nu^{2})(\nu-3ix-i\sqrt{m^{2}-\nu^{2}}) \right. \nonumber \\
& \left. +3x(2ix-\nu)(\sqrt{m^{2}-\nu^{2}}+x) \right] - \frac{C_{2}}{6m^{2}x}\exp\left(\frac{i\nu+\sqrt{m^{2}-\nu^{2}}}{x}\right) \times \nonumber \\
&\left[  (m^{2}-\nu^{2})(\nu-3ix+i\sqrt{m^{2}-\nu^{2}}) +3x(2ix-\nu)(\sqrt{m^{2}-\nu^{2}}-x)\right] \nonumber \\
&+ \frac{6C_{4}}{x^{2}}(i\nu+2x)-C_{6}(2i\nu+x)-\frac{10C_{3}}{x}-C_{7}x^{2}\exp\left(\frac{2i\nu}{x}\right)+C_{8}
\end{align}
Again, the trivial equation (\ref{eq:ltrivial}) is identically satisfied. The constraints (\ref{eq:uAc}) and (\ref{eq:uuc}) lead to

\begin{equation}
C_{8}-2\nu^{2}C_{6}=0\;,
\end{equation}

\begin{equation}
12i\nu C_{5}+6x\nu^{2}C_{6}+12(2C_{3}-3C_{4})-(3x-i\nu)C_{8}\;.
\end{equation}
We note that one can recover the static solutions by simply setting $\nu=0$ in the dynamical solution. With this in mind, we will only consider the dynamic case in the next section.

\subsubsection{Schwarzschild background}
\label{sec:schwarzschild}
When the background is Schwarzschild, we are not able to find analytical solutions in closed form. This is true even in general relativity for the case $M\neq 0$ and $\nu\neq 0$ \cite{Bishop:2004ug,cedeno:2015ucy}. In principle, one could write the solutions in terms of confluent hypergeometric functions or as a power series about the singular points of the concomitant ODEs. Here, we opt for the latter. The singular points of the ODEs (\ref{eq:rode}) and (\ref{eq:jode}) are as follows
\begin{align}
Regular:& \;x=\infty \qquad x=\frac{1}{2M} \\
Irregular:& \;x=0
\end{align}
In the following we compute series solutions about the regular singular point $x=1/2M$, corresponding to $r=2M$. We write $z=x-1/2M$ and expand the solutions about $z=0$, obtaining
\begin{align}
R  =& C_{1}\left[1+\frac{4M\left(2i\nu M+2m^{2}M^{2}+3\right)}{4i\nu M-1}z +\mathcal{O}\left(z^{2}\right)\right] \displaybreak[1]\\
\beta  =& C_{2} -C_{1}\left[\frac{4M^{2}m^{2}+8i\nu M+5}{12m^{2}\left(4i\nu M-1\right)} \right. +\nonumber \\
& \left. \frac{4M^{4}m^{4}+8M^{3}m^{2}i\nu-8\nu^{2}M^{2}+12m^{2}M^{2}+23i\nu M +6}{3m^{2}(4i\nu M-1)(2i\nu M-1)}z+\mathcal{O}\left(z^{2}\right) \right] \displaybreak[1]\\
J  =& C_{3}+C_{4}z+C_{5}\frac{z^{2}}{2}\left[1+\frac{8M(i\nu M+3)}{3\left(4i\nu M - 3 \right)}z +\mathcal{O}\left(z^{2}\right) \right]\displaybreak[1]\\
U  =&  \frac{C_{2}(2Mz+1)}{M} -i\nu C_{3}-3C_{4}(8M^{3}z^{3}+20z^{2}M^{2}+2i\nu M+14Mz+3) \nonumber \\
& -\frac{C_{5}(2i\nu M-1)}{8M^{3}}\left[1+4Mz+8M^{2}z^{2}+\mathcal{O}\left(z^{3}\right) \right]+\frac{2Mz+1}{12Mm^{2}}R\displaybreak[1]\\
w  =& C_{6}+\frac{40M^{2}z C_{2}}{2Mz+1}-C_{3}\left[\frac{2Mz+1+2M i\nu(Mz+1)}{(2Mz+1)^{2}} \right]48M^{2}z \nonumber \\
& + C_{4}\left[\frac{8 i\nu M (Mz+1)+4Mz(Mz+3)+5}{(2Mz+1)^{2}}\right]6Mz \nonumber \\
& + C_{1}\left[\frac{2M^{2}(16M^{3}m^{2}i\nu+16\nu^{2}M^{2}+36i\nu M-1 )}{3m^{2}(4 i\nu M-1)}z+\mathcal{O}(z^{2}) \right] \nonumber \\
& - C_{5}(2i\nu M-1)\left[6z-12Mz^{2}+32M^{2}z^{3} +\mathcal{O}(z^{4})\right]
\end{align}
This time, the trivial equation becomes a series in $z$, and is identically satisfied order by order. The constraints (\ref{eq:uAc}) and (\ref{eq:uuc}) respectively become

\begin{eqnarray}
3M^{2}m^{2}C_{6} &=& M^{2}(4Mi\nu+1)C_{1}-36M^{2}m^{2}C_{2}+72M^{2}m^{2}(Mi\nu+1)C_{3} \nonumber \\
&& -9Mm^{2}(2Mi\nu+3)C_{4}+6m^{2}(3i\nu M+2\nu^{2}M^{2}-1)C_{5}
\end{eqnarray}

\begin{eqnarray}
0 &=& 48M^{2}C_{2}-72M^{2}C_{3}+12M (2Mi\nu+3)C_{4}\nonumber \\
&& +\frac{16i\nu^{3}M^{3}-24\nu^{2}M^{2}-38i\nu M+15}{Mi\nu+2}C_{5} 
\end{eqnarray}
For the irregular singular point $x=0$, it is still possible to obtain a series solution for $J$ \cite{Bishop:2004ug}. However the same procedure does not work for $R$ (\ref{eq:rode}), hence a solution for the other quantities is not possible. In any case, standard methods for obtaining series solutions are not guaranteed to work for irregular singular points.

\section{Concluding remarks}
\label{sec:concluding_remarks}

In this work, we have presented a Characteristic formulation for metric $f(R)$ gravity. We have cast the full non-linear system both in tensorial form using the language of \cite{winicour:1983,winicour:1984} and also in the \textit{eth} formalism \cite{1967JMP.....8.2155G,Gomez:1996ge} that is commonly used in numerical relativity codes. The non-linear equations assume a simple structure as can be seen from \S\ref{sec:field_eqs} and \S\ref{sec:eth}, with $f(R)$ modifications encoded in the variables $M_{\beta}$, $M_{Q}$, $M_{W}$, and $M_{J}$. This makes it straightforward to modify existing codes that were originally built for general relativity to include terms arising from $f(R)$ gravity. 

A numerical implementation of the equations presented in this work will pave a way for Cauchy Characteristic Extraction methods in modified gravity. The recent detections of gravitational waves \cite{PhysRevLett.116.061102,Abbott:2016nmj} has opened up the possibility of constraining modified theories of gravity with gravitational wave data. This topic has revived some interest in the characterisation of gravitational radiation in $f(R)$ gravity theories \cite{PhysRevD.93.124071,Myung:2016zdl}. On the mathematical side, we have not addressed the Well-posedness of the timelike-null cone problem, upon which CCE is based. Interestingly, this is still an open question, even in general relativity. However, there has been encouraging results \cite{rendall:1990,Kreiss:2010ex}.

The linearised solutions presented in \S\ref{sec:minkowski_bg} will serve as testbed solutions for validating numerical codes. Another interesting area of application is in the linearised description of the binary black hole problem \cite{Bishop:2011iu,cedeno:2015mha,cedeno:2016cxx}. A potential application for this scenario is in the context of waveform extraction. Generally, one needs initial data on the null cone in some far field region exterior to a timelike worldtube. In this case, a linearised solution for the binary black hole problem presents a consistent approximation to the initial data \cite{Bishop:2011iu}. On the other hand, the series solutions in \S\ref{sec:schwarzschild} are somewhat of limited use as testbed solutions. This is largely due to their finite radius of convergence. However, they may still find analytical use in the study of gravitational wave scattering off a Schwarzschild black hole, which is a topic of broad interest see \cite{Sibandze:2016agp} and references therein.

Finally, we note that, in principle, applications of the Characteristic formulation of the field equations can go beyond numerical simulations. For example, one could use the formulation as an analytical tool to investigate various aspects of spherically symmetric solutions and their perturbations in $f(R)$ gravity \cite{Multamaki:2006zb,Capozziello:2007id,Nzioki:2009av,Clifton:2006ug,Canate:2015dda}. Using the Characteristic formulation in this way will allow for a transparent interpretation and generalisation of analytical results by using ready-built Characteristic codes. It would also be of interest to pursue comparisons with the covariant $1+1+2$ semi-tetrad formalism \cite{Clarkson:2002jz}. 

\section{Acknowledgements}
The author is grateful to Nigel Bishop and Obinna Umeh for valuable discussions and/or comments on earlier versions of this manuscript. The author also acknowledges financial support from the University of Cape Town Launching Grant programme.

\appendix
\section{Christoffel symbols}
\label{sec:chr_sym}
In the following we present the Christoffel symbols for the Bondi-Sachs metric (\ref{eq:bs_metric})
\begin{align}
\Gamma^{u}_{\phantom{u} AB} =&\; \sfrac{1}{2}e^{-2\beta}\pp_{r}\left(r^{2}h_{AB} \right)\\
\Gamma^{r}_{\phantom{r}rr} =&\; 2\pp_{r}\beta\displaybreak[1]\\
\Gamma^{A}_{\phantom{A}rB} =&\; \sfrac{1}{2}\left(r^{-2}h^{AC}\right)\pp_{r}(r^{2}h_{CB})\\
\Gamma^{u}_{\phantom{u}uu} =&\; -\sfrac{1}{2}e^{-2\beta}\left[2\pp_{u}(-e^{2\beta})+\pp_{r}\left(r^{-1}Ve^{2\beta} \right)-\pp_{r}\left(r^{2}h_{AB}U^{A}U^{B} \right)\right]\displaybreak[1]\\
\Gamma^{r}_{\phantom{r}rA} =&\; \sfrac{1}{2}e^{-2\beta}\left[\pp_{A}\left(e^{2\beta} \right)  +r^{2}h_{AB}\pp_{r}\left(U^{B} \right)\right]\displaybreak[1]\\
\Gamma^{r}_{\phantom{r}ru} =&\; -\sfrac{1}{2}e^{-2\beta}\left[\pp_{r}\left(-r^{-1}Ve^{2\beta} + r^{2}h_{AB}U^{A}U^{B}\right) \right] \nonumber \\
& -\sfrac{1}{2}e^{-2\beta}U^{A}\left[\pp_{r}\left(-r^{2}h_{AB}U^{B}\right) + \pp_{A}\left(e^{2\beta} \right) \right] \displaybreak[1]\\
\Gamma^{u}_{\phantom{u}uA} =&\; \sfrac{1}{2}e^{-2\beta}\left[\pp_{A}\left(e^{2\beta} \right)-U^{B}\pp_{r}\left(r^{2}h_{AB} \right)-r^{2}h_{AB}\pp_{r}\left(U^{B} \right) \right] \displaybreak[1]\\
\Gamma^{r}_{\phantom{r}uu } =&\; -\sfrac{1}{2}e^{-2\beta}\pp_{u}\left(-e^{2\beta}r^{-1}V+r^{2}h_{AB}U^{A}U^{B} \right) \nonumber \displaybreak[1]\\
& -\sfrac{1}{2}e^{-2\beta}r^{-1}V\left[2\pp_{u} \left(e^{-2\beta} \right)+ \pp_{r}\left(-e^{2\beta}r^{-1}V+r^{2}h_{AB}U^{A}U^{B}  \right)\right] \nonumber \displaybreak[1]\\
& +\sfrac{1}{2}e^{-2\beta}U^{A}\left[2\pp_{u}\left(r^{2}h_{AB}U^{B} \right) +\pp_{A}\left(-e^{2\beta}r^{-1}V +r^{2}h_{CD}U^{C}U^{D} \right) \right]  \displaybreak[1]\\
\Gamma^{r}_{\phantom{r}uA} =&\; -\sfrac{1}{2}e^{-2\beta}\left[\pp_{A}\left(-e^{2\beta}r^{-1}V + r^{2}h_{AB}U^{A}U^{B} \right) \right]\nonumber \displaybreak[1]\\
& - \sfrac{1}{2}r^{-1}Ve^{-2\beta}\left[\pp_{A}\left(e^{2\beta} \right) -\pp_{r}\left(r^{2}h_{AB}U^{B} \right)\right] \nonumber \displaybreak[1]\\
& - \sfrac{1}{2}e^{-2\beta}U^{B}\left[\pp_{u}\left(r^{2}h_{AB} \right)- \pp_{A}\left(r^{2}h_{CB}U^{C} \right) + \pp_{B}\left(r^{2}h_{AC}U^{C} \right)\right]\displaybreak[1]\\
\Gamma^{r}_{\phantom{r}AB} =&\; \sfrac{1}{2}e^{-2\beta}\left[2\pp_{A}\left(r^{2}h_{BC}U^{C} \right)+ \pp_{u} \left(r^{2}h_{AB} \right) \right] \nonumber \\
& -\sfrac{1}{2}r^{-1}Ve^{-2\beta}\left[\pp_{r}\left(r^{2}h_{AB} \right) \right]  - r^{2}e^{-2\beta}U_{D}{}^{(2)}\Gamma^{D}_{\phantom{D}AB} \\
\Gamma^{A}_{\phantom{A}uu} =&\; \sfrac{1}{2}e^{-2\beta}U^{A}\left[2\pp_{u}\left( e^{2\beta}\right) + \pp_{r}\left(-e^{2\beta}r^{-1}V+r^{2}h_{CD}U^{C}U^{D} \right) \right] \nonumber \\
& -\sfrac{1}{2}r^{-2}h^{AB} \left[2\pp_{u} \left(r^{2}h_{CB}U^{C} \right)+ \pp_{B} \left( -e^{2\beta}r^{-1}V + r^{2}h_{CD}U^{C}U^{D}\right)\right] \\
\Gamma^{A}_{\phantom{A}ur} =&\; \sfrac{1}{2}r^{-2}h^{AC} \left[\pp_{r}\left(-r^{2}h_{CD}U^{D} \right)+\pp_{C}\left(e^{2\beta} \right) \right]\\
\Gamma^{A}_{\phantom{A}Bu} =&\; \sfrac{1}{2}e^{-2\beta}U^{A}\left[\pp_{B}\left(e^{2\beta} \right) - \pp_{r}\left(r^{2}h_{CB}U^{C} \right)\right] \nonumber \\
& +\sfrac{1}{2}r^{-2}h^{AC}\left[\pp_{B} \left(-r^{2}h_{CD}U^{D} \right) + \pp_{u}\left(r^{2}h_{BC} \right) + \pp_{C}\left(r^{2}h_{BD}U^{D} \right)\right] \\
\Gamma^{A}_{\phantom{A}BC} =&\; \sfrac{1}{2}e^{-2\beta}U^{A}\left[\pp_{r}\left(r^{2}h_{BC} \right) \right] + {}^{(2)}\Gamma^{A}_{\phantom{A}BC}
\end{align}

In the above, ${}^{(2)}\Gamma^{A}_{\phantom{A}BC}$ represents the Christoffel symbols of the $2-$metric $h_{AB}$.

\end{document}